\documentclass[12pt,a4paper,twoside]{article}

\usepackage{graphicx}           	
\usepackage{caption}            	

\usepackage{natbib}			

\newcommand{\ignore}[1]{}   
\newcommand{\0}[1]{\phantom{0}}    

\begin{document}

\centerline{\large\bf EAS Tycho Brahe Prize Lecture 2011}
\centerline{ (to appear in Astron Astrophys Rev, 2011)}
\vspace{15pt}
\centerline{\large\bf Hipparcos: a Retrospective}
\vspace{10pt}
\centerline{Michael Perryman, School of Physics, University of Bristol}

\vspace{20pt}
\section*{Abstract}
The Hipparcos satellite was launched in 1989. It was the first, and remains to date the only, attempt at performing large-scale astrometric measurements from space. Hipparcos marked a fundamentally new approach to the field of astrometry, revolutionising our knowledge of the positions, distances, and space motions of the stars in the solar neighbourhood. In this retrospective, I look back at the processes which led to the mission's acceptance, provide a short summary of the underlying measurement principles and the experiment's scientific achievements, and a conclude with a brief summary of its principal legacy - the Gaia mission.


\vspace{10pt}
\noindent
Received: 9 September 2011

\section*{Introduction}
The Hipparcos space astrometry mission was accepted within the European Space Agency's science programme in 1980. It was duly launched in 1989 and, despite being stranded in a problematic elliptical geostationary transfer orbit, rather than its intended geostationary orbit, successfully operated until 1993. The final Hipparcos Catalogue of nearly 120\,000 stars, and the Tycho Catalogue of more than a million stars, were published in 1997. The enormous advance in accuracy of positions, parallaxes, and proper motions, and the accompanying elimination of systematic errors at levels well below 1~milliarcsec, ensured that the Hipparcos results would go on to have the substantial scientific impact  that had underpinned its acceptance almost two decades previously. 

I was ESA's project scientist for Hipparcos, from the time of its acceptance in 1980 until its completion in 1997. In this retrospective, I will present a brief history of its inclusion within Europe's space science programme, some key features of its operational principles, and a few highlights of its subsequent scientific results. All of these aspects are presented in further detail in my popular account of the Hipparcos project (Perryman, 2010).

I will also show how its scientific impact, and its efficient execution, led to the adoption of a much more ambitious space astrometry mission within Europe, Gaia.

\section{The Context}
For centuries astrometry, the accurate measurement of star positions, was the centrepiece of astronomical research, underpinning mankind's early appreciation of the nature of our Galaxy and its constituent stars. 

This was possible because accurate positional measurements yield not only a star's instantaneous location on the sky but, more importantly, and as derived from numerous measurements made over time, the star's motion through space, and it's distance from Earth (from the star's parallax). These ingredients represent the observational foundation for deriving the kinematic and dynamic structure of the Galaxy based on the former, and the stars' fundamental physical and evolutionary properties derived directly from the latter. The first measurements of trigonometric parallax in the 1830s led to a revolution in establishing the distance scale of the Universe, and in divining the physical properties of the stars. In the early 1900s, the improved knowledge of stellar distances became of considerable importance, and many new telescope initiatives were established over the following half century to respond to the challenge.

By the second half of the twentieth century, however, the steady advance in the accuracy of stellar positions was running into a number of essentially insurmountable barriers. Progress in telescopes and their instruments seen over the previous two or three centuries was running out of steam. Limited improvement in measuring star positions, in turn, obstructed further progress in fixing star distances and studying their space motions. For once, telescope size or optical quality were no longer the limiting factors. After two millennia of hard-won improvements, human ingenuity appeared to be finally barred by Nature's innate complexity. 

The biggest problem was the bending and twinkling effects of the atmosphere, condemning star images to their unpredictable wobbling dance. New thin-mirror technologies were having some success in correcting effects over small angles, but all attempts to determine large angles across the sky failed miserably. In addition, there were the variations in telescope alignment as the mountain-top observatories went through their day and night cycles of warming and cooling. The flexing of telescopes under their own weight as the supporting structures were steered to observe different parts of the sky added other distortions.

Yet another complication was that any telescope on Earth can observe only part of the sky at any one time: a telescope in the northern hemisphere only ever sees the northern skies. Even so, it still requires the best part of a year to elapse for the entire region to be observable by night. A grid of star positions spanning the entire sky could only be constructed from a vast network of thousands of geometrical triangulations from separate telescopes observing different portions of the sky at different times. However, between the various observations which had to be patched together, all of the star images had moved, all but chaotically, by the tiny amounts which were to be probed.

Like an ancient cartographic survey of the Earth made with primitive surveying instruments, the result of centuries of effort -- even into the 1980s -- was a map of the sky of sorts, but one which was distorted and warped. At accuracies below a second of arc, it was unreliable. Star positions were plagued by errors which could not be unravelled. Their space motions were, in consequence, of variable and sometimes questionable quality. More importantly, distances remained largely unknown, the signatures of their tiny parallaxes buried under a shroud of error-prone measurements imposed by the flickering atmosphere. A fundamentally new approach to measuring star positions was required.

\section{From Proposal to Acceptance}

The idea of dedicating an orbiting satellite to astrometry was, in hindsight, a master stroke of instrumental creativity. The basic concept, which I will outline below, was first formally presented to other scientists in the mid-1960s by the then 61-year old French astronomer Pierre Lacroute, although the idea of astrometric detection of binary stars and even exoplanets from space had been aired by Paul Couteau and Jean-Claude Pecker in a restricted bulletin of the Nice Observatory a couple of years before that. Until then space science, still very much in its youth, had been somewhat the preserve of magnetospheric experts studying the region of the Earth's environment controlled by its magnetic field, discovered by Explorer--1 in 1958. X-ray astronomers, meanwhile, were following up their discovery of the first cosmic X-ray source in 1962. It seems even more remarkable in hindsight that the push for astrometry, such a specialised goal in space science, should have followed so closely, within just a decade, of the first ever artificial satellite to orbit the Earth, the Soviet Union's Sputnik~1 in 1957.

But the idea of dedicating an expensive space platform to measure star positions was, for a number of years, neither enthusiastically received nor widely embraced. Beyond a limited group of its active exponents, the proposal probably appeared to many to be a misdirection of the limited opportunities of space funding.

Essentially, in the 1970s, astrometry found itself the victim of its own technical difficulties. Despite the efforts of brilliant instrumentalists, substantial progress had been slow because the problems were so forbidding. As a result, exciting new scientific results flowing from their work had largely dried up, and new creative minds were ill-inclined to enter the field. From the outside, the discipline probably appeared to be one which had run its course. Wide support for the funding of an expensive space mission would be all the more difficult to engender.

In its original conceptual design, Hipparcos was eventually rejected by the French national space agency, CNES, which had made the first assessment of its technical difficulties in the form outlined by Lacroute, and decided that it was technically too complex, and as a result excessively risky. But after submission to ESA, new minds took up and developed the challenge. Technical design innovations led by Erik H{\o}g from Copenhagen, fundamental advances in the concepts for the analysis of its data stream by Lennart Lindegren from Lund, and a concerted effort to formulate, explain and focus the interests of the wider astronomical community, notably by Catherine Turon from Meudon, Jean Kovalevsky from Grasse, and Pier Luigi Bernacca from Asiago, led to growing support for the mission, and a deeper appreciation of its fundamental importance.

The passage of Hipparcos through the various ESA advisory committees in the late 1970s and the first six months of 1980 was, however, not a particularly smooth one. As for all big projects competing at this level, different scientific opinions, conflicting national priorities, and different vested interests, made for difficult and sometimes heated exchanges within the relevant committees and advisory groups. At its meeting on 8--9~July 1980, the high-level ESA Science Programme Committee eventually decided in favour of the Giotto mission to Comet Halley as its highest priority, with Hipparcos to follow, the two missions squeezed into a challenging financial envelope. More details of these difficult and sometimes tense deliberations and decisions are given in my own popular account mentioned above.

The impressively-researched authorised history of ESA over the years 1958--1987 (by Krige, Russo, and Sebesta; and published by ESA Publications in 2000) sums it up perceptively after chronicling several similarly difficult choices over nearly two decades: {\it ``Choosing a big scientific project is also a matter of confrontation among scientists involved in the decision-making process: members of advisory committees or national delegations, government advisors, and policy makers. At each stage of the process, the traditional ties of cooperation, fellowship and solidarity that characterise the scientific community are strained by the emergence of national interests, disciplinary competitions, personal ambitions, career expectations, and personal relationships. When only one or two big projects can be started every three or four years the stakes are high and scientific objectivity is often a luxury. When making a choice entails some kind of painful discrimination, personal prestige, diplomatic talent, and personal or professional links can play a decisive role.''}

\section{The Operational Principles}

\begin{figure*}[t]
\centering
\includegraphics[width=0.86\linewidth]{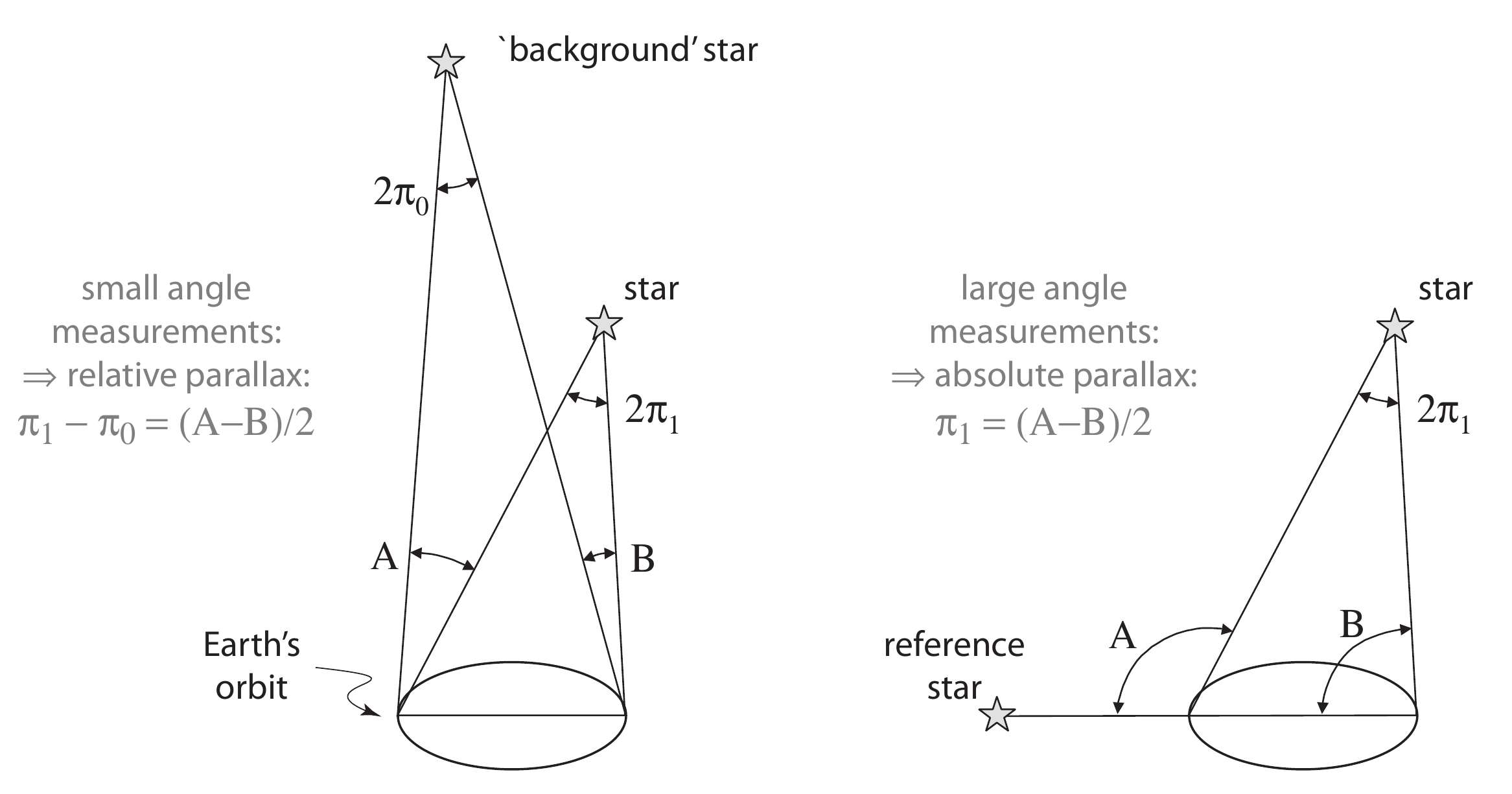}
\caption{\footnotesize The principle of absolute parallax determination.  Left: the measurement of the (small) angles $A,B$ only allows determination of the relative parallax $\pi_1-\pi_0=(A-B)/2$.  Right: in contrast, measurement of the large angles allows determination of the absolute parallax $\pi_1=(A-B)/2$, independent of the distance to the reference star (courtesy Lennart Lindegren).
\label{fig:lindegren05-04-ai-bw}
}
\end{figure*}

Placing a carefully designed optical telescope above the Earth's atmosphere was a necessary but insufficient element of the overall astrometric concept. A central feature were the two fields of view which looked out in widely-separated directions, combining the two viewing directions onto a common focal surface. This allowed a rigid reference to be set up which spanned the entire celestial sphere. More important, it provided the basis for the measurement of {\it absolute\/} rather than {\it relative\/} parallaxes (Figure~\ref{fig:lindegren05-04-ai-bw}). The separation angle, of a little more than $58^\circ$, was carefully chosen to optimise the stability of the positional reconstruction around the celestial sphere, simultaneously observing stars with very different parallax factors. [The split `beam-combining' mirror, at the heart of the Hipparcos payload, was one of its many challenging components, and the one that caused most concern during CNES's evaluation of the project. The flight spare of this 30~cm complex mirror is on loan to, and on display at, the National Maritime Museum at the Greenwich Observatory, UK.]

\begin{figure*}[t]
\centering
\includegraphics[width=0.5\linewidth]{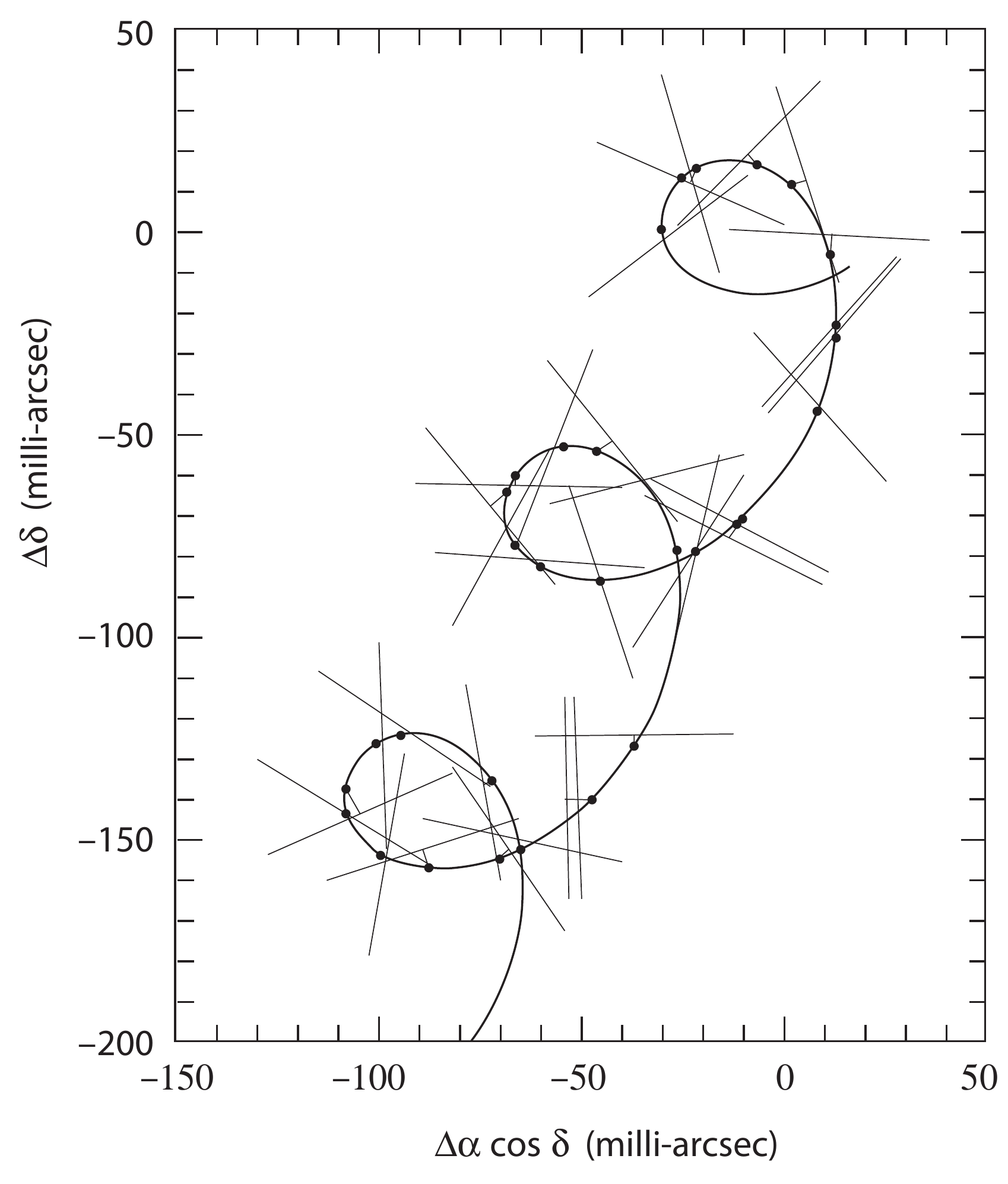}
\caption{\footnotesize The path on the sky of one of the Hipparcos stars, over the three-year observations.  Each straight line indicates the observed position of the star at a particular epoch: because the measurement is one-dimensional, the precise location along this position line is undetermined by the observation.  The curve is the modelled stellar path fitted to all the measurements. The inferred position at each epoch is indicated by a small solid circle, and the residual by a short line joining it to the corresponding position line. The amplitude of the oscillatory motion gives the star's parallax, with the linear component representing the star's proper motion. The intermediate astrometric data allow the quality of the model fitting to be assessed, and possibly refined. 
\label{fig:catvol1-2-8-1-ai}
}
\end{figure*}

As the satellite spun slowly around an axis orthogonal to these two viewing directions, with the spin axis itself precessing around the Sun direction in a carefully optimised manner, covering the sky but maintaining a constant (thermally stable) inclination to the Sun, the data stream sent to the ground contained the elements of a celestial jigsaw of a vast number of star positions, spread out over more than three years, encoding the tiny shifts of each star due to their proper motion and parallax (Figure~\ref{fig:catvol1-2-8-1-ai}). The ground processing was a huge task, with separate teams undertaking analysis of the main data channel and the Tycho data stream from the satellite's star mappers. The reductions were executed independently, but coordinated carefully, ensuring rigour and confidence in the final data products. Computers across Europe crunched through the calculations for six years until the final results were in a form that could be published. 

Eventually, Hipparcos produced not only the most accurate star catalogue ever compiled, but one with uniform and carefully quantified systematics. It delivered the largest improvement factor in astrometric accuracy in history, even surpassing -- for example -- that provided by the hugely ambitious observations of Tycho Brahe. And, for the community waiting for its scientific harvest, the final results were delivered, validated and documented, just three years after the end of the satellite observations.

Only through a considerable effort by many extremely talented and committed scientists, over a significant part of their professional careers, were the Hipparcos results finally achieved. In spite of such a large structure, with many important contributions by many people, primary responsibilities fell to relatively few: for the scientific content of the final catalogue, in terms of the stars carefully selected pre-launch for measurement, to Catherine Turon; and for the execution of the data analysis, teams under the lead of Erik H{\o}g, Jean Kovalevsky, and Lennart Lindegren. For various details of the instrument design, Erik H{\o}g's insight was indispensable. And for a significant number of specific mathematical optimisations of the system, ranging from the basic `three-step' numerical data reduction on ground, optimisation of the scanning law, the separation angle of the two viewing directions within the instrument, of the detailed reflective Schmidt optical design, and various other key aspects, Lennart Lindegren of the Lund Observatory in Sweden was central to the overall design, ensuring rigour in the instrumental optimisation, and in the data analysis. Many others played key parts in the photometry, in the double star processing, in establishing a quasi-inertial reference frame link, in the processing of solar system objects, and so on.

It is important to recognise also the skills and commitments of the industrial teams, so central to the technical realisation of such a complex satellite: considerable technical ingenuity was required to solve many engineering issues, such as the accurate thermal control, the very complex attitude control, data formatting and transmission on-board, and the overall stability of the optical system. Integration and testing posed many further challenges, under the relentless dual pressures of the impending launch and a fixed budget. The launch authority, Arianespace, held responsibility for placing the immeasurably valuable and highly-delicate satellite into orbit, and the ESA operations team at ESOC, Germany, under the operations manager Dietmar Heger, ran the operations for its immensely difficult four-year lifetime. 

After launch in August 1989, the project suffered a major setback: the apogee boost motor failed to ignite, and the satellite was left in an unintended elliptical transfer orbit, at the mercy of a twice-daily immersion in the Van~Allen radiation belts. The mission seemed to be all-but-over before the observations had even started, and most had given up hope of retrieving even a small fraction of the mission's original goals. I was delegated as the overall mission project manager for the operational phase, as well as its project scientist, and with the support of the scientific and operations teams behind me, took the project slowly from recovery and on to its eventual success. The numerous difficulties that we faced during this phase, arising from the unintended orbit, and complicated by the network of four ground-stations that had to be scrambled to collect the data, are also recounted in more detail in my popular account (Perryman, 2010).

\section{The Hipparcos Project Organisation}

I will not expand here on further details of the 8-year development phase of Hipparcos, between the start of the detailed design in 1981 and launch in 1989, except to stress one very important point: that the scientific goal of 2~milliarcsec accuracy for 100\,000 stars, which was at the very foundation of the mission's acceptance in 1980, and the ESA member states subscription to the project, was fully preserved throughout the technical challenges that the mission faced during its detailed design (Phase~B) and construction (Phase~C/D). 

In a few words, and with the perspective of hindsight, I attribute this eventual success to a combination of key factors:
(a)~a properly-conducted feasibility study before the mission's acceptance in 1980 -- both scientific and technical -- and its associated costing; 
(b)~committed management by the two successive ESA Hipparcos project managers, Franco Emiliani and Hamid Hassan;
(c)~excellent industrial teams leading and contributing to the technical design and manufacture, led by EADS Astrium in France (formerly Matra Marconi, led by Michel Bouffard) and Alenia Spazio in Italy (led by Bruno Strim);
(d)~a highly committed and disciplined scientific effort, led by Catherine Turon (as leader of the INCA Input Catalogue Consortium), and Erik H{\o}g, Jean Kovalevsky, and Lennart Lindegren (as leaders of the three data analysis teams);
(e)~a 12-strong and highly focused Hipparcos Science Team, led by myself, and responsible for all scientific aspects of the mission;
(f)~a careful binding of all of these groups into a single committed structure, with mutual respect for the very different roles, expertise, and responsibilities of each.

The final results, even despite the wrong orbit, surpassed the demands and expectations placed on the satellite at the time of its acceptance by ESA's Science Programme Committee in 1980. What had been promised in 1980 was a catalogue of 100\,000 stars, with positions, parallaxes and annual proper motions, each with a mean accuracy of 2~milliarcsec. What was delivered in 1997 was a catalogue of just less than 120\,00 stars with corresponding accuracies of below 1~milliarcsec, and even somewhat better than that after Floor van Leeuwen's 2007 reanalysis of the final stages of the reduction, with improved computational facilities, and a deeper insight into the satellite's attitude evolution. Accompanying catalogues of state-of-the-art photometry, and double and multiple star parameters, were published in parallel. On top of this, Erik H{\o}g's brainchild, the Tycho Catalogue, delivered lower accuracy but still of enormous positional value for more than a million stars and, in 2000, a reanalysis in combination with the 100-year old Astrographic Catalogue, gave improved positions and proper motions for a remarkable 2.5~million stars. 

The Millennium Star Atlas, a beautiful and informative set of star charts, was compiled and published under a splendid collaboration with a team from Sky Publishing in the US led by Roger Sinnott. It set the standard for a number of other celestial atlases making use of the Hipparcos data which have since followed.

The final catalogues, and the first scientific results from them, were presented at an international symposium in Venice in 1997. Many of those who had been involved for many years in the project, and others from beyond Europe, gathered in the beautiful island location of San Giorgio, Venice, to inspect its content.

Distributed to the world's scientific libraries in 1997, the Hipparcos results are freely accessible through data centres, and repeatedly used for countless investigations. As a catalogue of star positions as they were arranged in the sky around 1990, it maps a configuration which will never be seen again. As such, astrometry bestows a rare treasure in science; it provides a unique snapshot of the Universe at one moment, preserved for all time. It may diminish in relevance as future star catalogues from space build upon it, but its historical value will persist indefinitely.

{\it ``Altogether thirty years elapsed before our work was completed''}, said Jean Kovalevsky at an ESA award ceremony in 1999, at which Kovalevsky, H{\o}g, Lindegren and Turon received the ESA Director of Science's medal, at a ceremony of the Science Programme Committee in Bern, a decade after launch. {\it ``For individuals involved from the beginning, it was an extraordinary commitment within a human lifetime. Yet thirty years was a short time in the history of science to achieve a revolution that has affected every branch of astronomy''}. 

As Roger--Maurice Bonnet enthused during the presentation: {\it ``As team leaders, our medalists were responsible for the largest computing task in the history of astronomy. ESA says thank you to them and the many other scientists who devoted twenty or thirty years of their working lives to making Hipparcos a success.''}

\section{The Scientific Harvest}

Several thousand scientific papers based on the catalogues have appeared since 1997. An extensive compilation of these various scientific results, with references, is given in my book review of the Hipparcos scientific results (Perryman, 2009).

\begin{figure*}[t]
\centering
\includegraphics[width=0.7\linewidth]{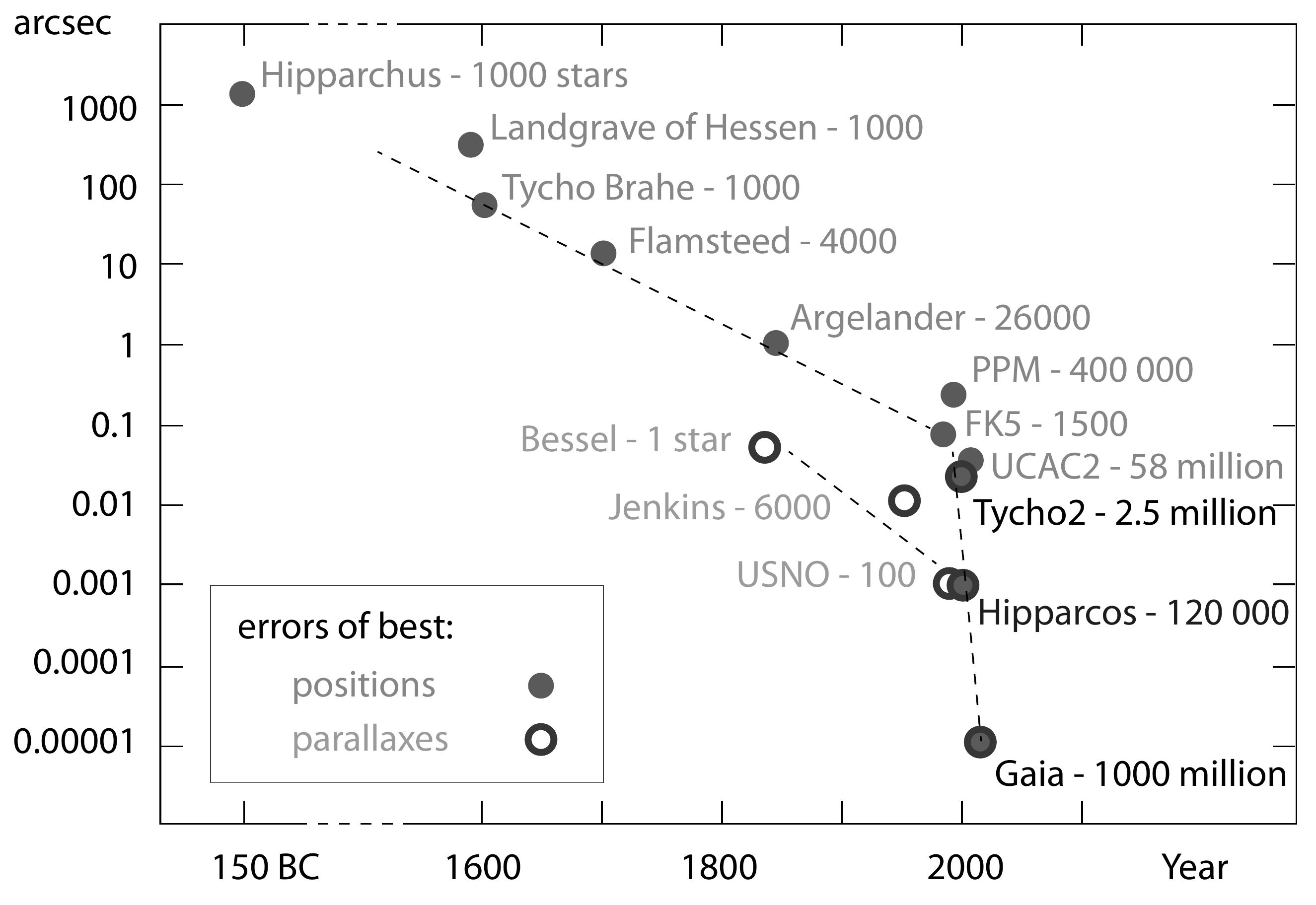}
\caption{\footnotesize The accuracy of the Hipparcos and Tycho Catalogue astrometry, placed in their historical context. Separate accuracy tracks are shown for positions and parallaxes. Hipparcos provided not only the most accurate stellar positions, but also the largest improvement factor ever achieved (courtesy Erik H{\o}g).
\label{fig:astrometric-accuracy-hoeg-ai-bw}
}
\end{figure*}

As a consequence of the improvement in astrometric accuracy (illustrated in a historical context in Figure~\ref{fig:astrometric-accuracy-hoeg-ai-bw}), the Hipparcos astrometric results impact a very broad range of astronomical research, which I will classify into three major themes:

\paragraph{Provision of an Accurate Reference Frame} 
This has allowed the consistent and rigorous re-reduction of a wide variety of historical and present-day astrometric measurements. The former category include Schmidt plate surveys, meridian circle observations, 150~years of Earth-orientation measurements, and re-analysis of the 100-year old Astrographic Catalogue (and the associated Carte du Ciel). The Astrographic Catalogue data, in particular, have yielded a dense reference framework reduced to the Hipparcos reference system propagated back to the early 1900s. Combined with the dense framework of 2.5~million star mapper measurements from the satellite, this has yielded the high-accuracy long-term proper motions of the Tycho~2 Catalogue.

The dense network of the Tycho~2 Catalogue has, in turn, provided the reference system for the reduction of current state-of-the-art ground-based survey data: thus the dense UCAC~3 and USNO~B2 catalogues are now provided on the same reference system, and the same is true for surveys such as SDSS and 2MASS.  Other observations specifically reduced to the Hipparcos system are the SuperCOSMOS Sky Survey, major historical photographic surveys such as the AGK2 and the CPC2, and the more recent proper motion programmes in the northern and southern hemispheres, NPM and SPM. Proper motion surveys have been rejuvinated by the availability of an accurate optical reference frame, and amongst them are the revised NLTT (Luyten Two-Tenths) survey, and the L\'epine--Shara proper motion surveys (north and south).  Many other proper motion compilations have been generated based on the Hipparcos reference system, in turn yielding large data sets valuable for open cluster surveys, common-proper motion surveys, etc.

The detection and characterisation of double and multiple stars has been revolutionised by Hipparcos: in addition to binaries detected by the satellite, many others have been revealed through the difference between the Hipparcos (short-term) proper motion, and the long-term photocentric motion of long-period binary stars (referred to as $\Delta\mu$~binaries).  New binary systems have been followed up through speckle and long-baseline optical interferometry from ground, through a re-analysis of the Hipparcos Intermediate Astrometric Data, or through a combined analysis of astrometric and ground-based radial velocity data.  Other binaries have been discovered as common-proper motion systems in catalogues reaching fainter limiting magnitudes. Various research papers have together revised the analysis of more than 15\,000 Hipparcos binary systems, providing new orbital solutions, or characterising systems which were classified by Hipparcos as suspected double, acceleration solutions, or stochastic (`failed') solutions.

Other studies since 1997 have together presented radial velocities for more than 17\,000 Hipparcos stars since the catalogue publication, not counting two papers presenting some 20\,000 radial velocities from the Coravel data base, and more than 70\,000 RAVE measurements including some Tycho stars.  These radial velocity measurements are of considerable importance for determining the three-dimensional space motions, as well as further detecting and characterising the properties of binary stars.

More astrophysically, studies have been made of wide binaries, and their use as tracers of the mass concentrations during their Galactic orbits, and according to population. Numerous papers deal with improved mass estimates from the spectroscopic eclipsing systems, important individual systems such as the Cepheid-binary Polaris, the enigmatic Arcturus, and favourable systems for detailed astrophysical investigations such as V1061~Cyg, HIP~50796, the mercury-manganese star $\phi$~Her, and the spectroscopic binary HR~6046. A number of papers have determined the statistical distributions of periods and eclipse depths for eclipsing binaries, and others have addressed their important application in determining the radiative flux and temperature scales. Studies of the distributions of detached and contact binaries (including W~UMa and symbiotic systems) have also been undertaken.

The accurate reference frame has in turn provided results in topics as diverse as the measurement of General Relativistic light bending; Solar System science, including mass determinations of minor planets; applications of occultations and appulses; studies of Earth rotation and Chandler Wobble over the last 100~years based on a re-analysis of data acquired over that period within the framework of studies of the Earth orientation; and consideration of non-precessional motion of the equinox.

\begin{figure*}[t]
\centering
\includegraphics[width=0.95\linewidth]{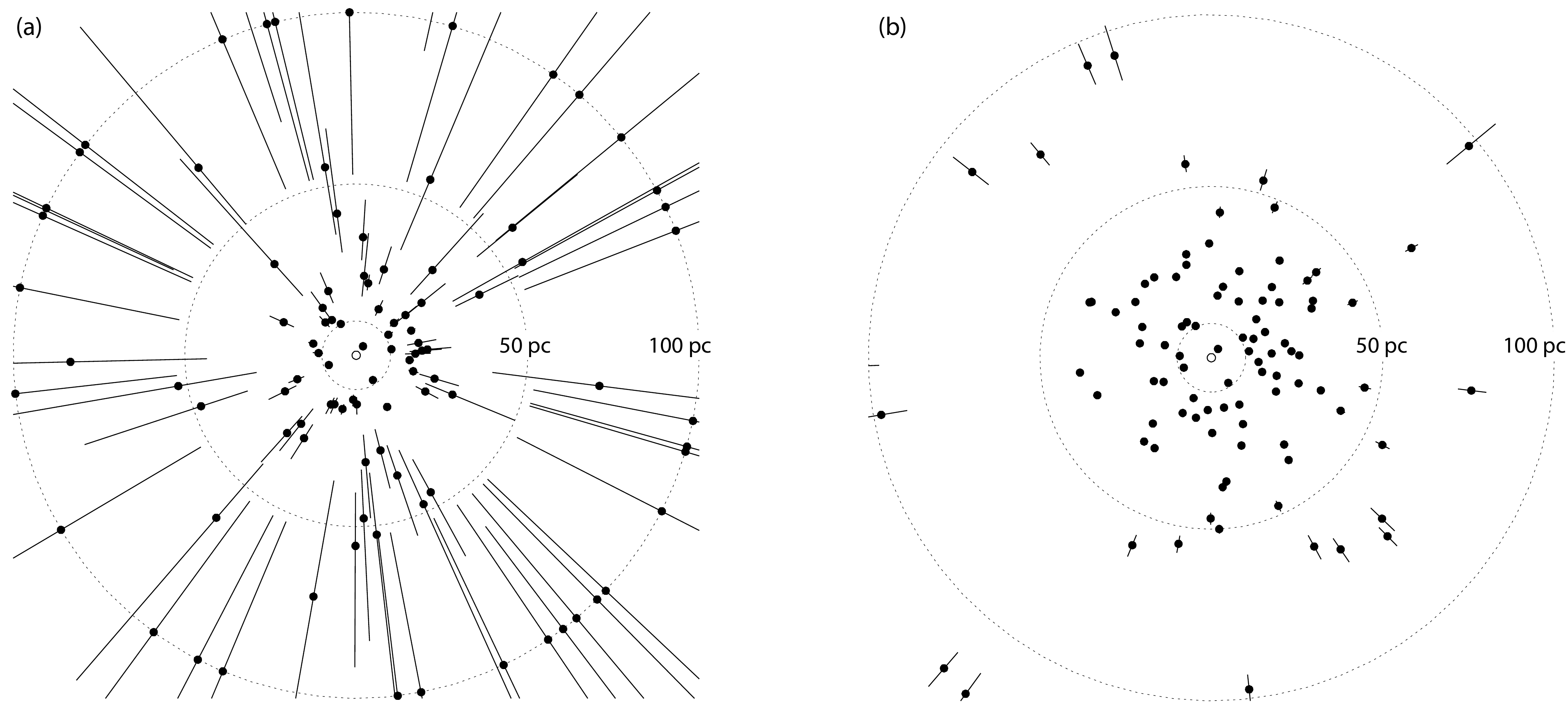}
\caption{\footnotesize Improvement in the knowledge of exoplanet host star distances by Hipparcos. For the 100 brightest stars with exoplanets known from radial velocity measurements at the end of 2010 ($V<7.2$~mag), estimated distances and standard errors are shown from: (a)~the ground-based compilation of Van Altena et al.\ (General Catalogue of Trigonometric Stellar Parallaxes, 1995), and (b)~from Hipparcos. Azimuthal coordinates correspond to right ascension, independent of declination. Distances undetermined in~(a) are arbitrarily assigned a parallax of $10\pm9$\,mas. Hipparcos substantially improved both parallax standard errors, and their systematics.
\label{fig:perryman-distances-gcstp-hipparcos}
}
\end{figure*}

\paragraph{Constraints on Stellar Evolutionary Models}
The accurate distances and luminosities of 120\,000 stars has provided the most comprehensive and accurate data set relevant to stellar evolutionary modeling to date, providing new constraints on internal rotation, element diffusion, convective motions, and asteroseismology. Combined with theoretical models it yields evolutionary masses, radii, and stellar ages of large numbers and wide varieties of stars. A simple illustration of the improvement brought to stellar distances is given in Figure~\ref{fig:perryman-distances-gcstp-hipparcos}, which shows the knowledge of some of the recently-discovered exoplanet host star distances pre- and post-Hipparcos.

A substantial number of papers have used the distance information to determine absolute magnitude as a function of spectral type, with new calibrations extending across the Hertzsprung--Russell diagram: for example for OB stars, AFGK~dwarfs, and GKM~giants, with due attention given to the now more-quantifiable effects of Malmquist and Lutz--Kelker biases.  Other luminosity calibrations have used spectral lines, including the Ca~{\footnotesize II}-based Wilson--Bappu effect, the equivalent width of O~{\footnotesize I}, and calibrations based on interstellar lines.

A considerable Hipparcos-based literature deals with all aspects of the basic `standard candles' and their revised luminosity calibration. Studies have investigated the Population~I distance indicators, notably the Mira variables, and the Cepheid variables (including the period--luminosity relation, and their luminosity calibration using trigonometric parallaxes, Baade--Wesselink pulsational method, main-sequence fitting, and the possible effects of binarity). The Cepheids are also targets for Galactic kinematic studies, tracing out the Galactic rotation, and also the motion perpendicular to the Galactic plane.

Hipparcos has revolutionised the use of red clump giants as distance indicators, by providing accurate luminosities of hundreds of nearby systems, in sufficient detail that metallicity and evolutionary effects can be disentangled, and the objects then used as single-step distance indicators to the Large and Small Magellanic Clouds, and the Galactic bulge.  Availability of these data has catalysed the parallel theoretical modeling of the clump giants.

For the Population~II distance indicators, a rather consistent picture has emerged in recent years based on subdwarf main sequence fitting, and on the various estimates of the horizontal branch and RR~Lyrae luminosities.

A number of different methods now provide distance estimates to the Large Magellanic Cloud, using both Population~I and Population~II tracers, including some not directly dependent on the Hipparcos results (such as the geometry of the SN~1987A light echo, orbital parallaxes of eclipsing binaries, globular cluster dynamics, and white dwarf cooling sequences). Together, a convincing consensus emerges, with a straight mean of several methods yielding a distance modulus of $(m-M)_0=18.49$~mag. Through the Cepheids, the Hipparcos data also provide good support for the value of the Hubble Constant $H_0=72\pm8$\,km\,s$^{-1}$\,Mpc$^{-1}$ as derived by the Hubble Space Telescope key project, and similar values derived by the Wilkinson Microwave Anisotropy Probe (WMAP), gravitational lensing experiments, and Sunyaev--Zel'dovich effect.

A wide range of other studies has made use of the Hipparcos data to provide constraints on stellar structure and evolution. Improvements have followed in terms of effective temperatures, metallicities, and surface gravities. Bolometric corrections for the {\it Hp\/} photometric band have opened the way for new and improved studies of the observational versus theoretical Hertzsprung--Russell diagram.

Many stellar evolutionary models have been developed and refined over the last few years, and Hipparcos provides an extensive testing ground for their validation and their astrophysical interpretation: these include specific models for pre- and post-main sequence phases, and models which have progressively introduced effects such as convective overshooting, gravitational settling, rotation effects, binary tidal evolution, radiative acceleration, and effects of $\alpha$-element abundance variations.  These models have been applied to the understanding of the Hertzsprung--Russell diagram for nearby stars, the reality of the B\"ohm-Vitense gaps, the zero-age main sequence, the subdwarf main sequence, and the properties of later stages of evolution: the subgiant, first ascent and asymptotic giant branch, the horizontal branch, and the effects of dredge-up, and mass-loss.  Studies of elemental abundance variations include the age--metallicity relation in the solar neighbourhood, and various questions related to particular elemental abundances such as lithium and helium. Other studies have characterised and interpreted effects of stellar rotation, surface magnetic fields, and observational consequences of asteroseismology, notably for solar-like objects, the high-amplitude $\delta$~Scuti radial pulsators, the $\beta$~Cephei variables, and the rapidly-oscillating Ap stars.

Many studies have focused on the pre-main sequence stars, both the (lower-mass) T~Tauri and the (higher-mass) Herbig Ae/Be stars, correlating their observational dependencies on rotation, X-ray emission, etc.  The understanding of Be stars, chemically-peculiar stars, X-ray emitters, and Wolf--Rayet stars have all been substantially effected by the Hipparcos data. Kinematic studies of runaway stars, produced either by supernova explosions or dynamical cluster ejection, have revealed many interesting properties of runaway stars, also connected with the problem of (young) B~stars found far from the Galactic plane.

Dynamical orbits within the Galaxy have been calculated for planetary nebulae and, perhaps surprisingly given their large distances, for globular clusters, where the provision of a reference frame at the 1~milliarcsec accuracy level has allowed determination of their space motions and, through the use of a suitable Galactic potential, their Galactic orbits, with some interesting implications for Galactic structure, cluster disruption, and Galaxy formation.

One of the most curious of the Hipparcos results in this area is the improved determination of the empirical mass--radius relation of white dwarfs.  At least three such objects appear to be too dense to be explicable in terms of carbon or oxygen cores, while iron cores seem difficult to generate from evolutionary models.  `Strange matter' cores have been postulated, and studied by a number of groups.

\paragraph{Galactic Structure and Dynamics}
The distances and uniform space motions have provided a substantial advance in understanding of the detailed kinematic and dynamical structure of the solar neighbourhood, ranging from the presence and evolution of clusters, associations and moving groups, the presence of resonance motions due to the Galaxy's central bar and spiral arms, the parameters describing Galactic rotation, the height of the Sun above the Galactic mid-plane, the motions of the thin disk, thick disk and halo populations, and the evidence for halo accretion.

Many attempts have been made to further understand and characterise the solar motion based on Hipparcos data, and to redefine the large-scale properties of Galactic rotation in the solar neighbourhood. The latter has been traditionally described in terms of the Oort constants, but it is now evident that such a formulation is quite unsatisfactory in terms of describing the detailed local stellar kinematics.  Attempts have been made to re-cast the problem into the 9-component tensor treatment of the Ogorodnikov--Milne formulation, analogous to the treatment of a viscous and compressible fluid by Stokes more than 150 years ago. The results of several such investigations have proved perplexing. The most recent and innovative approach has been a kinematic analysis based on vectorial harmonics, in which the velocity field is described in terms of (some unexpected) `electric' and `magnetic' harmonics. They reveal the warp at the same time, but in an opposite sense to the vector field expected from a stationary warp.

Kinematic analyses have tackled the issues of the mass density in the solar neighbourhood, and the associated force-law perpendicular to the plane, the $K_z$ relation. Estimates of the resulting vertical oscillation frequency in the Galaxy of around 80~Myr have been linked to cratering periodicities in the Earth's geological records. Related topics include studies of nearby stars, the stellar escape velocity, the associated initial mass function, and the star formation rate over the history of the Galaxy. Dynamical studies of the bar, of the spiral arms, and of the stellar warp, have all benefited. Studies of the baryon halo of the Galaxy have refined its mass and extent, its rotation, shape, and velocity dispersion, and have provided compelling evidence for its formation in terms of halo substructure, some of which is considered to be infalling, accreting material, still ongoing today.

New techniques have been developed and refined to search for phase-space structure (i.e.\ structure in positional and velocity space): these include convergent-point analysis, the `spaghetti' method, global convergence mapping, epicycle correction, and orbital back-tracking. An extensive literature has resulted on many aspects of the Hyades, the Pleiades, and other nearby open clusters, comprehensive searches for new clusters, and their application to problems as diverse as interstellar reddening determination, correlation with the nearby spiral arms, and the age dependence of their vertical distribution within the Galaxy: one surprising result is that this can be used to place constraints on the degree of convective overshooting by matching stellar evolutionary ages with cluster distances from the Galactic plane.

In addition to studying and characterising open clusters, and young nearby associations of recent star formation, the Hipparcos data have revealed a wealth of structure in the nearby velocity distribution which is being interpreted in terms of open cluster evaporation, resonant motions due to the central Galactic bar, scattering from nearby spiral arms, and the effects of young nearby kinematic groups, with several having been discovered from the Hipparcos data in the past few years.

The Hipparcos stars have been used as important (distance) tracers, determining the extent of the local `bubble', itself perhaps the result of one or more nearby supernova explosions in the last 5~Myr. The interstellar medium morphology, extinction and reddening, grey extinction, polarisation of star light, and the interstellar radiation field, have all been constrained by these new distance estimates of the Hipparcos stars.

\paragraph{Other Applications}

Superficially it may seem surprising that the Hipparcos Catalogue has been used for a number of studies related to the Earth's climate. Studies of the passage of nearby stars and their possible interaction with the Oort Cloud have identified stars which came close to the Sun in the geologically recent past, and others which will do so in the relatively near future. Analysis of the Sun's Galactic orbit, and its resulting passage through the spiral arms, favour a particular spiral arm pattern speed in order to place the Sun within these arms during extended deep glaciation epochs in the distant past. In this model, climatic variations are explained as resulting from an enhanced cosmic-ray flux in the Earth's atmosphere, leading to cloud condensation and a consequent lowering of temperature.  A study of the Maunder Minimum, a period between 1645--1715 coinciding with the coldest excursion of the `Little Ice Age', and a period of great hardship in Europe, was interpreted in the context of the number of solar-type stars out to 50--80~pc showing correspondingly decreased surface activity.  Several studies have used the accurate distance data, accompanied by stellar evolutionary models, in an attempt to identify `solar twins' (stars which most closely resemble the Sun in all its characteristics, and which may be the optimum targets for searches for life in the future), and `solar analogues' (stars which will resemble the Sun at some past or future epoch, and which therefore offer the best prospects for studying the Sun at different evolutionary stages).

Many studies have used the accurate photometric data: as part of the construction of absolute or bolometric magnitudes, for their uniform colour indices, and for their extensive epoch photometry, which itself has been used for all sorts of variability analyses, including the rotation of minor planets, the study of eclipsing binaries, the complex pulsational properties of Cepheids, Mira variables, $\delta$~Scuti, slowly-pulsating B~stars, and many others.

In addition to all of these, the Hipparcos and Tycho Catalogues are now routinely used to point ground-based telescopes, navigate space missions, drive public planetaria, and provide search lists for programmes such as exoplanet surveys; one study has even shown how positions of nearby stars at the milliarcsec level can be used to optimise search strategies for extraterrestrial intelligence.

\section{The Hipparcos Legacy: Gaia}

Hipparcos left another important legacy, beyond that of the scientific advance that it provided. The results, and the acquisition of the technical and scientific knowledge it entailed, inspired a number of new satellite studies intended to advance the field further. However, the German national project DIVA, intended as a rapid follow-up, met its demise as the ESA Gaia project gained momentum. The somewhat comparable accuracy Japanese mission Nano-Jasmine, remains under development, with launch foreseen in 2012. The US Naval Observatory advanced two or three different project designs to carry the field further, with their JMAPS project currently still under development. All of these national efforts have underlined that, 30~years after the Hipparcos mission was accepted, it is still extremely challenging to make astrometric measurements, even at the milliarcsec level, from space.

Nevertheless, ESA's Gaia project arose out of the general perception that the principles established by Hipparcos could be carried much further, and very rapidly, with the expertise that had been fostered by the enormous Hipparcos effort. The first ideas for a European-level follow-up mission emerged around 1995, and were carried forward by a large team of enthusiastic scientists, their sights set on the remarkable results that a further huge leap in accuracy promised. 

This next instrumental advance will follow similar principles, but with both scientific ambition and the experiment itself scaled up to reflect twenty years of progress in astronomy and technology, leapfrogging the Hipparcos accuracy by a factor one hundred. It will feature a far bigger lightweight telescope, built from the highly stable ceramic silicon carbide. Like a massive video camera, a carpet of CCD sensors almost a square meter in area will record the millions of star images that pass across it as a new orbiting satellite once more scans the heavens.  A powerful on-board processor will handle a vast cascade of image manipulations before the information stream is despatched to Earth. Its data rate will be more than a hundred times that of its predecessor. The satellite's orbit will be rather different---far from Earth, one and half million kilometers away at the Sun--Earth Lagrange point. 

This next leap in ambition promises a scientific harvest which dwarfs that of Hipparcos. Its colossal survey of more than a thousand million suns will provide a defining census of around one per cent of our Galaxy's entire stellar population, pin-pointing them in space right across its vast expanses. Unimaginable numbers of stellar motions will reveal many more details of the vastly complex motions at play within our Galaxy. It will provide insights ranging from new tests of general relativity to stringent limits on the variation of fundamental physical constants. Planets circling other stars will appear in their thousands from their tiny wobbling motions, identifying candidate systems for the burgeoning discipline of exobiology. Tens of thousands of asteroids will be measured. 

After five years of study, Gaia was accepted by the ESA Science Programme Committee in 2000, just as Hipparcos had been twenty years before. Its goals were to survey every star in the sky to 20~mag, and to deliver an astrometric catalogue at the level of 10~$\mu$as at 15~mag, degrading towards fainter magnitudes in proportion to photon statistics. To try to convey something of this instrumental challenge, 10~$\mu$as corresponds to the angle subtended by one Bohr radius viewed from a distance of one metre. Current plans are for a launch in 2013.

\vspace{10pt}\noindent
I chaired the Gaia Science Team, and coordinated its scientific development, from 1995 to 2008. I stood down from my position as Gaia project scientist in 2008, with concerns about the successive descoping imposed on the project after its selection, and -- not unrelated -- its technical management. There are plateaus in accuracy for any experiment at which certain scientific objectives are, or are not achieved. The successive descopes, from the 10~$\mu$as accuracy at acceptance in 2000, to 15~$\mu$as in 2002, 20~$\mu$as in 2004, and then to 25~$\mu$as in 2006, meant that a number of key scientific objectives in areas such as Galactic structure, near-Earth object detection, and exoplanetary science, were substantially degraded. Compared to the 30\,000 Jupiter-mass systems detectable to some 200~parsec at 10~$\mu$as, for example, the numbers will drop by a factor of $(2.5)^3$, i.e.\ to around 2000, corresponding to the volume of space probed at a given accuracy. Unfortunately not only detections, but direct mass determinations (rather than $M\sin i$), and the characterisation of resonance systems and relative orbital inclinations, all of fundamental importance for constraining formation and evolution models, will be correspondingly degraded, to levels significantly below those identified at the time of selection. 

While Gaia remains an extraordinary advance over the first ever space astrometry mission, descoping seems to have become a routine management tool {\it post-selection}, rather than being reserved as an exceptional response to exceptional circumstances.

\vspace{10pt}\noindent
In closing, I emphasise the obvious, that there is still much about the Universe to discover and to understand. Astronomy in general, and astrometry in particular, have an immensely exciting future that we are well-positioned to explore further. Space missions are inherently difficult and problematic, but we do continue to live in very privileged times when large projects are still funded, and in which fertile scientific and technical minds can pool their resources, leading to great advances in knowledge that would have appeared improbable even one or two decades before. 

\vspace{10pt}\noindent
Significant scientific and technological progress requires considerable vision, good judgement, strong leadership, great organisation, and a variety of other ingredients.  Along with some good fortune, underpinned by much hard work, even the occasional `miracles' can happen, allowing substantial successes even when the situation might have appeared desperate.

\section*{Acknowledgments} 

Astronomy progresses substantially through instrumental advances, as well as observation and theory, and the Tycho Brahe Prize of the European Astronomical Society has been established, relatively recently, to recognise such contributions.

I am grateful to the EAS, and to its President Thierry Courvoisier, for their recognition of the importance of the Hipparcos mission within the exciting and extensive overall panorama of astronomical instrumentation, and my role within the project. I thank those of my colleagues who supported my nomination, and I pass on my gratitude and respect to the very large number of scientists, engineers, managers, politicians, and funding authorities who all played a part in turning Pierre Lacroute's vision of almost half a century ago, into a remarkable reality. I also thank the Klaus Tschira Foundation for supporting this prize.

\vfill
\noindent
This paper will be appear in Astron Astrophys Rev, 2011.
The final publication is available at springerlink.com.


\begin{thebibliography}{2}
\providecommand{\natexlab}[1]{#1}
\itemsep 1pt

\bibitem[\protect\astroncite{{Perryman}}{2009}]{2009aaat.book.....P}
{Perryman} MAC, 2009, \emph{{Astronomical Applications of Astrometry: Ten Years
  of Exploitation of the Hipparcos Satellite Data}}. Cambridge University Press

\bibitem[\protect\astroncite{{Perryman}}{2010}]{2010mhgs.book.....P}
{Perryman} MAC, 2010, \emph{{The Making of History's Greatest Star Map}}. Springer-Verlag, Berlin Heidelberg

\end{thebibliography}

\end{document}